\documentclass[conference]{IEEEtran}
\IEEEoverridecommandlockouts
\usepackage{cite}
\usepackage{enumitem}
\usepackage[numbers,sort&compress]{natbib}
\usepackage{booktabs}
\usepackage{tabularx}
\usepackage{float}
\usepackage{hyperref}
\usepackage{amsmath}
\usepackage{extdash}
\newcommand{\cmmnt}[1]{}
\usepackage{amsmath,amssymb,amsfonts}
\usepackage{algorithmic}
\usepackage{graphicx}
\usepackage{textcomp}
\usepackage{xcolor}
\def\BibTeX{{\rm B\kern-.05em{\sc i\kern-.025em b}\kern-.08em
    T\kern-.1667em\lower.7ex\hbox{E}\kern-.125emX}}
\begin{document}

\title{DELRec: Distilling Sequential Pattern to Enhance LLMs-based Sequential Recommendation}

\author{
\IEEEauthorblockN{Haoyi Zhang\IEEEauthorrefmark{0\text{1},\dag}, 
Guohao Sun\IEEEauthorrefmark{0\text{1},\dag}, 
Jinhu Lu\IEEEauthorrefmark{0\text{1}}, 
Guanfeng Liu\IEEEauthorrefmark{0\text{2}},
Xiu Susie Fang\IEEEauthorrefmark{0\text{1},\ddag}
\thanks{$^\dag$Equal Contribution}
\thanks{$^\ddag$Corresponding Author}
}\IEEEauthorblockA{\IEEEauthorrefmark{0\text{1}}Donghua University, Shanghai, 201620, China}\IEEEauthorblockA{\IEEEauthorrefmark{0\text{2}}Macquarie University, Sydney, NSW 2109, Australia}\IEEEauthorblockA{2232980@mail.dhu.edu.cn, ghsun@dhu.edu.cn, 1209126@mail.dhu.edu.cn,\\ guanfeng.liu@mq.edu.au, xiu.fang@dhu.edu.cn}
}

\maketitle
\begin{abstract}
Sequential recommendation (SR) tasks aim to predict users' next interaction by learning their behavior sequence and capturing the connection between users' past interactions and their changing preferences. Conventional SR models often focus solely on capturing sequential patterns within the training data, neglecting the broader context and semantic information embedded in item titles from external sources. This limits their predictive power and adaptability. Large language models (LLMs) have recently shown promise in SR tasks due to their advanced understanding capabilities and strong generalization abilities. Researchers have attempted to enhance LLMs-based recommendation performance by incorporating information from conventional SR models. However, previous approaches have encountered problems such as 1) limited textual information leading to poor recommendation performance, 2) incomplete understanding and utilization of conventional SR model information by LLMs, and 3) excessive complexity and low interpretability of LLMs-based methods.

To improve the performance of LLMs-based SR, we propose a novel framework, \underline{D}istilling Sequential Pattern to \underline{E}nhance \underline{L}LMs-based Sequential \underline{Rec}ommendation (DELRec), which aims to extract knowledge from conventional SR models and enable LLMs to easily comprehend and utilize the extracted knowledge for more effective SRs. DELRec consists of two main stages: 1) \textit{Distill Pattern from Conventional SR Models}, focusing on extracting behavioral patterns exhibited by conventional SR models using soft prompts through two well-designed strategies; 2) \textit{LLMs-based Sequential Recommendation}, aiming to fine-tune LLMs to effectively use the distilled auxiliary information to perform SR tasks. Extensive experimental results conducted on four real datasets validate the effectiveness of the DELRec framework.
\end{abstract}


\section{INTRODUCTION}
\label{three_para}
Sequential recommendation \cite{sequential_challenges,survey_sr,sr_survey_2} predicts users' next interaction by learning their behavior sequence. SR aims to improve the accuracy of recommendations by understanding and modeling the relationship between users' interaction history and their evolving preferences. Conventional SR models \cite{rss_conv} only capture sequential patterns within training data, often overlooking the broader context and semantic information embedded in item titles that can be obtained from external sources. These limitations restrict their predictive ability and adaptability to constantly changing scenarios.

Large language models have recently shown pro\-mise in SR tasks due to their advanced comprehension abilities and powerful generalization capabilities \cite{recommendation_as_instruction, p5}. As LLMs are trained on vast datasets containing abundant information, including inherent item features and details, they can infer user preferences and predict future actions \cite{learning_phrase_rnn_enco} by leveraging LLMs' understanding of item attributes and inferring based on world knowledge. However, using LLMs directly as sequential recommenders \cite{tallrec} can pose certain problems. For instance, due to a lack of domain-specific expertise \cite{align_know} in the recommendation or an incomplete understanding of the recommendation patterns in SR tasks, LLMs often perform poorly when directly used as recommenders. 

Thus, researchers have previously proposed integrating auxiliary information from conventional SR models into LLMs. These approaches aim to provide information from conventional SR models to LLMs to enhance the accuracy of SR tasks conducted by LLMs. The LLMs-based SR integrating conventional SR models can be roughly categorized into three paradigms. These three paradigms are illustrated in Figure \ref{figure intro1}, and the detailed introduction and issues caused by them are summarized below:\\[0.25em]
\textbf{Providing LLMs with textual information extracted from conventional SR models}. This paradigm typically involves incorporating the textual information extracted from conventional SR models into the prompt \cite{sr_prompt_issue}. However, this paradigm often suffers from poor recommendation performance due to the limited information provided by the prompt. One fundamental reason is that textual information is often insufficient for accurately and comprehensively describing conventional SR models' specific recommendation information and behavior patterns.\\[0.25em]
\textbf{Providing LLMs with embeddings from conventional SR models}. This paradigm typically merges embeddings from conventional SR models with a prompt before inputting them into LLMs to generate item recommendations. This paradigm uses embeddings encoded by conventional SR models as auxiliary information for the recommendation process provided to LLMs and often involves a projector to align the dimensions of conventional SR model's embeddings with the language space of LLMs. However, due to poor projector design or changes in embedding dimensions \cite{adapter_problem_1_kellmrec} that result in information loss, LLMs may not fully comprehend the meanings conveyed by these embeddings.\\[0.25em]
\textbf{Combining embeddings from LLMs and conventional SR models}. Instead of using LLMs as recommenders, this paradigm utilizes LLMs' encoding and representation capabilities. It typically involves utilizing LLMs to encode a given text or sequence and simultaneously employing conventional models to obtain item or user encodings. Subsequently, these two types of embeddings are combined and processed in various ways to generate recommendation scores for items. Although this paradigm enables the integration of information from both conventional SR models and LLMs, it also introduces challenges in comprehending and interpreting recommendations. This may potentially undermine some key advantages of using LLMs for recommendations, such as their simplicity and interpretability.

\begin{figure}[htbp]
\vspace{-2mm}
  \centering
  \includegraphics[width=0.95\linewidth]{fig_intro_m1.pdf}
  \caption{Demonstration of three paradigms of the integration of conventional SR models with LLMs-based SR.}
  \label{figure intro1}
  \vspace{-1mm}
\end{figure}

To tackle the aforementioned problems, we propose \underline{D}istill\-ing Sequential Pattern to \underline{E}nhance \underline{L}LMs-based Sequential \underline{Rec}ommenda\-tion (DELRec), which aims to distill the behavioral patterns of conventional SR models and empower LLMs to easily comprehend and leverage this supplementary information for more effective SRs. The paradigm of DELRec is roughly shown in Figure \ref{Figure del}, and it contains two stages:\\[0.25em]
\textbf{Distill Pattern from Conventional SR Models}. Rather than inputting encoded information from conventional SR models or LLMs as previous methods did, the first stage of DELRec (\textit{Distill Pattern from Conventional SR Models}) is inspired by knowledge distillation \cite{prompt_distill} techniques used in LLMs. The objective is to distill conventional SR models' recommendation patterns and information that are understandable to LLMs. This stage involves using LLMs to extract useful knowledge of conventional SR models into soft prompts \cite{soft_the}. Specifically, this stage consists of two key components: \textit{Temporal Analysis} (TA) and \textit{Recommendation Pattern Simulating} (RPS). Through \textit{Temporal Analysis}, LLMs are able to simulate conventional SR models for temporal analysis, thereby gaining temporal awareness. \textit{Recommendation Pattern Simulating} allows LLMs to learn the recommendation results of conventional SR models, enabling them to simulate the recommendation behavior patterns of conventional SR models. During this stage, LLMs conduct tasks of these two components concurrently through multi-task learning and adjust the parameters of the soft prompts. Consequently, LLMs are empowered to effectively comprehend and simulate the recommendation process employed by conventional SR models. This is a process of transforming the knowledge of conventional SR models into a form that LLMs can fully utilize.\\[0.25em]
\textbf{LLMs-based Sequential Recommendation}. After getting the distilled SR knowledge in the first stage for SR tasks, we propose \textit{LLMs-based Sequential Recommendation} for effectively instructing LLMs. Instead of using a projector for embedding mapping, we insert the learned soft prompts into the prompt to avoid information inefficiency or loss caused by using a projector to align embedding dimensions and then fine-tune the LLMs to adapt to the learning tasks that utilize auxiliary information.

\begin{figure} [htbp]
  \vspace{-3mm}
  \includegraphics[width=0.95\linewidth]{delrec_rough.pdf}
  \vspace{-2mm}
  \caption{Demonstration of the paradigm of DELRec.}
  \label{Figure del}
  \vspace{-7mm}
\end{figure}
\vspace{5mm}

The main contributions of our work are summarized as follows:
\begin{itemize}[leftmargin=*]
 
\item Proposing a novel framework, DELRec, to enhance the LLMs-based SR performance with conventional SR models.

\item Proposing two novel components in the first stage of DELRec to distill the information and behavior patterns of conventional SR models with soft prompts.

\item Designing an ingenious method in the second stage of DELRec to fine-tune LLMs, enabling them to utilize the auxiliary information and achieve more accurate SR tasks.

\item Conducting extensive experiments to demonstrate the effectiveness of DELRec.
\end{itemize}

\section{RELATED WORK}
This section presents a literature review on Conventional Sequential Recommendation and LLMs-based Sequential Recommendation.

\subsection{Conventional Sequential Recommendation}
In the field of SR, recognizing the sequential pattern of user interactions is crucial for predicting their next preference.

For early research of conventional SR models, Markov chain models are often used for SR tasks. It mainly relies on probability theory and statistics to model state transitions in sequential data. FPMC \cite{FPMC} combines the ideas of Markov chains and matrix factorization. Fossil \cite{Fossil} integrates an item similarity model with a high-order Markov chain, leveraging both the item matrix and user historical behavior.

With the continuous development of deep learning, more and more conventional SR models also employ various deep learning techniques such as  Recurrent Neural Networks (RNNs), Convolutional Neural Networks (CNNs), or Transformers to identify sequential patterns in user interactions. 

GRU4Rec \cite{GRU4Rec} uses Gated Recurrent Units (GRUs), a type of RNN, to analyze users' interactions. It predicts users' next actions by effectively capturing their long-term dependencies during a session. Caser \cite{Caser} uses CNNs to process multi-dimensional interaction data, where interactions between users and items are encoded as a two-dimensional matrix, with each element representing a specific interaction. SASRec \cite{SASRec} utilizes an attention mechanism to process interaction data, where the behavior of users is encoded as vectors and serves as inputs for the attention model. TIGER \cite{TIGER} builds semantically meaningful codeword tuples as the semantic ID for each item, and trains a transformer-based model to predict the target item.

Although these methods are good at capturing sequential patterns, they fail to effectively utilize the semantic information contained within item titles.

\subsection {LLMs-based Sequential Recommendation}
\label{KDA_label}
With the ongoing development of LLMs, researchers are exploring methods to integrate conventional SR models with LLMs \cite{text_is,learning_vector,distill_matter} in order to enhance the performance of SR tasks. We illustrate the existing methods according to the three paradigms summarized in the introduction section.\\[0.25em]
\textbf{Providing LLMs with textual information extracted from conventional SR models}. PISA \cite{promptcast} takes numerical sequences as input and generates numerical outputs. It converts input and output into prompts and predicts sentence by sentence using a LLM. RecRanker \cite{recranker} uses importance-aware and clustering-based sampling to collect user data for training. It enhances performance by integrating recommendation results from conventional models into textual prompts. Llama4Rec \cite{Llama4Rec} proposes customized data augmentation and prompt augmentation strategies to enhance both conventional models and LLMs. It uses an adaptive aggregation module to combine the textual results of the two models into prompts, refining the final recommendation outcome. LLMSEQPROMPT \cite{LLMSeqSim} improves SR accuracy by injecting domain knowledge into the prompts of LLMs. It uses the session containing the item list as a prompt and the last item name as completion, and then fine-tunes the LLMs. LLM-TRSR \cite{LLM-TRSR} segments users' interactions and generates recurrent summaries using LLMs, then it builds prompts that contain users' preference summaries, recent interactions, and candidate information, then finally uses prompts to fine-tune LLMs. Despite using textual information from conventional SR models to prompt LLMs, the limited information provided by textual information leads to poor performance of LLMs-based SR.
\\[0.25em]
\textbf{Providing LLMs with embeddings from conventional SR models}. LLaRA \cite{llara} leverages multimodal mapping by inserting prompts with item embeddings encoded by conventional SR models, and then fine-tunes LLMs with item interaction relationships. LLMGR \cite{LLMGR} uses a tokenizer to extract text embeddings and conventional models to extract graph embeddings. It also develops auxiliary and main instructions to adjust task prompts and effectively combine text and graph embeddings, thereby enhancing the effectiveness of SR. RLMRec \cite{RLMRec} combines representation learning with LLMs to capture intricate semantic embeddings of user behaviors and preferences. RLMRec uses LLMs for user/item profiling to incorporate auxiliary textual signals, and it aligns these semantic embeddings with collaborative relational textual signals through cross-view alignment. LLM2BERT4Rec \cite{LLMSeqSim} initializes item embeddings in BERT4Rec \cite{bert4rec} with LLMs embeddings, and reduces the dimensionality of LLMs embeddings using PCA \cite{PCA} as projector before initialization. Then, it trains BERT4Rec with these embeddings using the BERT's \cite{bert} masked language model training protocol. However, these methods typically require a projector to align embedding dimensions and semantic space. Poorly designed projectors can lead to information loss or ineffective utilization by LLMs.\\[0.25em]
\textbf{Combining embeddings from LLMs and conventional SR models}. LLM-based Generation of Item-Description \cite{Item_bert} utilizes LLMs to generate the description and then employs BERT to encode the movie's information along with its original movie ID. The combined embeddings obtained from these two steps are then input into a conventional recommendation model in order to obtain the recommended item. LlamaRec \cite{llamarec} employs a conventional model to generate embeddings for item recall, passes these embeddings to LLMs, and ultimately utilizes a verbalizer to transform the output logits into a candidate probability distribution. RA-Rec \cite{RA-Rec} first converts the users' browsing history sequences into natural language, then uses prompt-tuning to guide LLMs to encode them to the user embeddings, and finally combines them with the item embedding extracted from SR models to obtain scores for the predicted items. LLMSEQSIM \cite{LLMSeqSim} uses LLMs to obtain item embeddings, and it calculates session embeddings by combining embeddings of preference items within the session. Then, it computes similarity to recommend the most similar item to the user. Relation-aware SR with Latent Relation Discovery (LRD) \cite{LRD} utilizes the LRD framework based on LLMs to predict the latent relationship between items, and it reconstructs item embeddings based on these relationships. Then, these latent relationships are integrated into the recommender models along with the embeddings extracted by conventional SR models. However, these methods may undermine key advantages of using LLMs for recommendations, such as simplicity and interpretability.

The proposed DELRec utilizes soft prompts to extract information and recommendation patterns from conventional SR models by carefully designing components. These extracted soft prompts are then inserted into the prompt, allowing for fine-tuning of LLMs to learn this auxiliary information. DELRec effectively addresses previous problems and improves the performance of LLMs-based SR.

\section{PRELIMINARY}

\subsection{Task Formulation} We consider a recommender system with a set of users $U$, where a user $u \in U$ has an interaction sequence that consists of a sequence of $n$ items $(I_1,I_2,...,I_n)$ in chronological order ($n$ can be different for different users). The SR task is defined as follows: given the user interaction sequence $I_{1:n-1} = (I_1, I_2,..., I_{n-1})$, a sequential recommender aims to predict the target item $I_n$ from a set of candidate items $C$, where candidate set that consists of $m$ items $(I_1, I_2,..., I_m)$ is typically selected from the entire item set $I$, where $m \ll |I|$. 

Unlike conventional SR models, we leverage LLMs to solve the recommendation task in an instruction-following paradigm. Specifically, for each user $u$, we construct a history prompt including the users' interactions $I_{1:n-1} = (I_1, I_2,..., I_{n-1})$, a candidate item prompt including the candidate items $C$, and soft prompts including the recommendation patterns and information of conventional SR models. The aforementioned prompts are concatenated along with an instruction that explicitly describes the recommendation task, forming the final prompt $P$ for LLMs. Finally, LLMs employ the prompt $P$ to predict the target item $I_n$.

\vspace{-1mm}

\subsection{Soft Prompt} Hard prompts, also known as discrete prompts, are composed of specific vocabulary; unlike hard prompts, soft prompts remove the constraint that prompt embeddings must correspond to natural language words \cite{construc__prompt}. Soft prompts can be adjusted according to downstream task training data, so they can provide LLMs with "\textit{only LLMs understand}" knowledge that is difficult or impossible for humans to describe in natural language. Although hard prompts usually correspond to natural language and are easily understood by humans, the purpose of prompt construction is to enable LLMs to perform tasks effectively, not for human consumption; additionally, during LLMs inference, both hard and soft prompts are ultimately passed to LLMs in the form of word embeddings, so it is not necessary to limit prompts to human-interpretable natural language \cite{softprompt_survey}. Hence, unlike the general prompt, we will insert a portion of soft prompts into the construction of our prompts. Formally, we denote soft prompts as $sp_j$ and hard prompts as $hp_i$, where $j$ is the index of soft prompts in the prompt, and $i$ is the index of hard prompts in the prompt. Our prompt $P$ is constructed by both hard and soft prompts: 
\begin{equation}
\begin{aligned}
P = \{ hp_1, hp_2, ..., sp_1, sp_2, ..., sp_k, ..., hp_{l-1}, hp_l \},
\end{aligned}
\end{equation}

\noindent where $k$ represents the number of soft prompts in the prompt $P$ and $l$ represents the number of hard prompts in the prompt $P$. Both hard and soft prompts in our prompt $P$ will also become word embeddings. However, unlike hard prompts corresponding to a fixed position in the language space, soft prompts will be processed into randomly initialized embeddings. As LLMs learn the target task, the position of the soft prompts in the language space will change:
{\small
\begin{equation}
\begin{aligned}
E_1 = \sum_{j=1}^{k}f_{iniz}( sp_j )
\end{aligned}
\end{equation}
}

\noindent where $E_1$ represents the corresponding embeddings directly initialized by the soft prompts, and $f_{iniz}$ indicates the process of randomly initializing to the same dimension as the word embeddings in the language space of LLMs.

\begin{figure*}
\vspace{-6mm}
  \centering
  \includegraphics[width=0.95\textwidth]{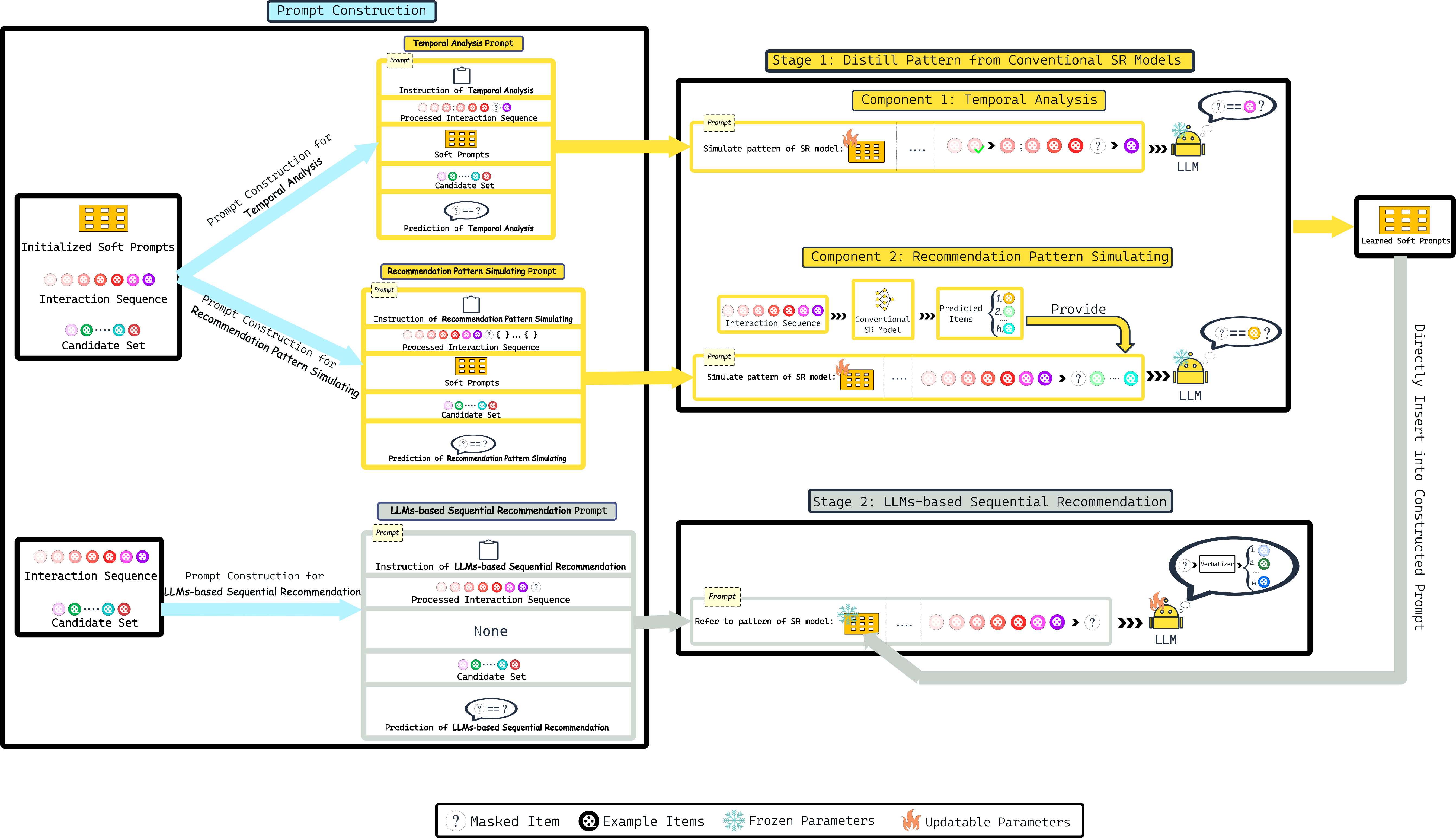}
  \caption{Illustrating the proposed DELRec. There are three parts in DELRec: \textit{Prompt Construction} aims to construct prompts for the next two stages to guide LLMs to perform different tasks. \textit{Stage 1: Distill Pattern from Conventional SR Models} involves using soft prompts to distill SR patterns from conventional SR models. This stage includes two components: \textit{Temporal Analysis} and \textit{Recommendation Pattern Simulating}. \textit{Temporal Analysis} focuses on performing time analysis on conventional SR models and providing similar time knowledge to LLMs; \textit{Recommendation Pattern Simulating} aims to simulate the recommendation results of conventional SR models, enabling soft prompts to distill similar recommendation knowledge. \textit{Stage 2: LLMs-based Sequential Recommendation} aims to insert distilled soft prompts into constructed prompt and fine-tune LLMs, enabling them to predict the ground truth.}
  \label{Figure 1}
  \vspace{-6mm}
\end{figure*}

\subsection{Parameter Efficient Fine-Tuning}The comprehensive fine-tuning of all parameters within LLMs demands considerable time and computational resources. To mitigate this issue, the approach of Parameter-Efficient Fine-Tuning (PEFT) concentrates on adjusting a minimal subset of parameters, thereby reducing computational demands while maintaining notable performance levels. An example of a PEFT method is AdaLoRA (Adaptive LoRA), which optimizes the number of trainable parameters for weight matrices and layers, unlike LoRA, which evenly distributes parameters across all modules. It allocates more parameters to important weight matrices and layers, while less important ones receive fewer parameters. The optimization goal for AdaLoRA is formulated as follows:
\begin{equation}
\begin{aligned}
\max_{\Theta} \sum_{(x,y) \in D}  \sum_{t=1}^{|y|} \log (P_{\Phi_0+\Delta\Phi_0(\Theta)}(y_t|x, y_{<t})),
\end{aligned}
\end{equation}

\noindent where AdaLoRA introduces the parameters \( \Theta \), which are smaller than the original LLM parameters \( \Phi_0 \).

\section{METHODOLOGY}
We propose DELRec framework to improve the performance of LLMs-based SR. The overview is depicted in Figure \ref{Figure 1}. 

In the beginning, DELRec constructs three distinct prompts for different tasks. In the first stage, DELRec updates soft prompts' parameters through two components, \textit{Temporal Analysis} and \textit{Recommendation Pattern Simulating}, to distill information and recommendation patterns from conventional SR models. In the second stage, DELRec integrates the distilled recommendation pattern into LLMs and fine-tunes them to improve performance in conducting more accurate SR tasks. 

\subsection{Prompt Construction}
\label{prompt_construction}

We construct two different prompts for the two components of the first stage, and construct one prompt for the second stage. We create three different prompts to help LLMs comprehend and execute the diverse tasks in DELRec. In the general template, all three prompts consist of instruction, processed interaction sequence, candidate set, soft prompts, and prediction. Details of these components are as follows:\\[0.25em]
\textbf{Instruction}. Each of these three prompts contains unique instructions that specify LLMs' tasks.\\[0.25em]
\textbf{Processed Interaction Sequence}. The interaction sequence changes based on the three distinct tasks, offering diverse contexts and information sources to the LLMs, which constitutes the most crucial part of the prompts.\\[0.25em]
\textbf{Candidate Set}. The candidate set consists of one item to be predicted and a series of other random items. This guides LLMs in selecting the target item from the set to ensure that non-existent items are not included in their output.\\[0.25em]
\textbf{Soft Prompts}. The Soft Prompts part evolves through different stages. For the first stage, initialized soft prompts are inserted into this part to distill information and recommendation patterns from conventional SR models. For the second stage, this part is initially set to none, but after the first stage, learned soft prompts are inserted to provide distilled information, enabling LLMs to make more accurate SRs.\\[0.25em]
\textbf{Prediction}. The content of the prediction includes masked parts that require the LLMs to predict. It aims to output what the masked item should be according to the instructions.

Specifically, for the task of the first component (\textit{Temporal Analysis}) in the first stage, the prompt we construct aims to let the LLMs analyze the temporal dynamics of the conventional SR models. This component involves updating the parameters of soft prompts, thereby allowing LLMs to simulate the recommendation patterns of conventional SR models at a process level. The detailed content will be introduced in Section \ref{temporal_analy}. For the task of the second component (\textit{Recommendation Pattern Simulating}) in the first stage, our constructed prompt is intended to facilitate LLMs in learning and simulating recommendation results from conventional SR models. It also updates soft prompt parameters, enabling LLMs to simulate recommendation patterns at a result level. Further details will be provided in Section \ref{recommendation_pattern}.

Regarding tasks in the second stage, our constructed prompt aims to guide LLMs towards considering previously learned soft prompts more as a reference. It is worth noting that, when constructing the prompt for the second stage, we first leave the places for soft prompts empty and insert the learned soft prompts after the first stage. Additionally, we fine-tune LLMs to let them utilize auxiliary information extracted from conventional SR models for predicting ground truth, ultimately enhancing their ability to perform more accurate SR. The detail will be introduced in Section \ref{LLMs-sr}.

It is crucial to highlight that when populating the prompts with information, unlike conventional SR tasks where IDs denote items, we represent all items in the prompts (candidate set, processed interaction sequence, prediction) using textual titles. This approach equips LLMs with richer intrinsic details about the items, enhancing their ability to execute the designated tasks effectively. Furthermore, we will incorporate specific names of the conventional SR models to harness the pre-existing knowledge of LLMs and facilitate a deeper understanding of these conventional SR models' recommendation patterns.

\subsection{Distill Pattern from Conventional SR Models}
The first stage of DELRec, \textit{Distill Pattern from Conventional SR Models}, uses soft prompts inserted into the constructed prompts to accurately capture the information and recommendation behavior patterns of conventional SR models for LLMs. We propose this stage to provide LLMs with information from conventional SR models to enhance the performance of LLMs as recommenders. Specifically, this stage consists of two components, namely \textit{Temporal Analysis} and \textit{Recommendation Pattern Simulating}.\\[0.25em]
\textbf{Temporal Analysis}. 
\label{temporal_analy}
Since SR tasks aim to recommend items that are temporally closer based on user interaction sequences, which exhibits strong temporal dynamics, it is crucial to perform a temporal analysis of conventional SR models and provide similar temporal knowledge to LLMs to better simulate conventional SR models' recommendation patterns.

Most conventional SR models ($e.g.$, GRU4Rec, SASRec) achieve temporal awareness by aggregating the features of items in the user interaction sequence to the most recent item in the sequence. We aim to enable LLMs to similarly recognize and learn the importance of "the most recent item," thereby acquiring relevant temporal knowledge. Therefore, our proposed strategy is to provide the interaction sequence and target item, and let the LLMs predict the most recent item in the sequence——a behavior we refer to as PMRI (Predicting Most Recent Item). 

\begin{figure}[htbp]
 \vspace{-3mm}
  \centering
  \includegraphics[width=0.95\linewidth]{temporal_analysis_prompt.pdf}
  \caption{Demonstration of the prompt for \textit{Temporal Analysis}.} 
  \label{Figure 4}
  \vspace{-2mm}
\end{figure}

Specifically, our strategy will allow LLMs to perform PMRI on the sequences, and we will also provide in-context learning (ICL) \cite{ICL} in an ingenious way to not only help LLMs enhance their learning efficiency and quality, but also increase LLMs' temporal awareness, our strategy is as follows:

Given the user interaction sequence  $I_{1:n-1} = (I_1, I_2, ..., I_{\alpha-1}, I_{\alpha}, ..., I_{n-2}, I_{n-1})$, we then inform LLMs that the $\alpha$-th item will be the next interaction item for the sequence of the first $k$ items $I_{1:\alpha-1}$, which takes the previous part of the sequence as ICL provided to the LLMs. Similarly, we take the last item $I_{n-1}$ as the next item for a sequence $I_{\alpha:n-2}$ and mask the second-to-last item $I_{n-2}$, allowing LLMs to predict the masked item $I_{n-2}$ and assign it as the label $y^0$ for this task. During the prediction process, we will use a simple verbalizer to effectively convert the output of the LLM head ($i.e.$, the output scores of all tokens) into ranking scores for all items. In the learning process, the parameters of the LLMs \( \Phi_0 \) are frozen, and only the parameters of the soft prompts \( \Phi \) are updated. Afterward, the soft prompts in the prompt will contain knowledge of the conventional SR models' aggregation of item features and knowledge similar to the temporal information of conventional SR models. The prompt of this component is shown in Figure \ref{Figure 4}. Formally, the learnable parameters of soft prompts \( \Phi \) are optimized by minimizing the loss function of \textit{Temporal Analysis} (TA):

\vspace{-1.5mm}
\begin{equation}
\begin{aligned}
 L_{TA} = \sum_{(x^0,y^0) \in D_0} -\log 
(P_{\Phi_0+\Phi}(y^0 | x^0)), 
\end{aligned}
\vspace{-1mm}
\end{equation}

\noindent where $D_0 = \{(x_i^0, y_i^0)\}_{i=1, ...,N}$ contains the prompt and masked item in the aforementioned.\\[0.25em]
\textbf{Recommendation Pattern Simulating}. 
\label{recommendation_pattern}
Besides \textit{Temporal Analysis}, it is also essential for LLMs to simulate conventional SR models in making similar recommendations, which enables the distillation from the recommendation knowledge of conventional SR models into soft prompts.

\begin{figure}[htbp]
 \vspace{-2mm}
  \centering
  \includegraphics[width=0.95\linewidth]{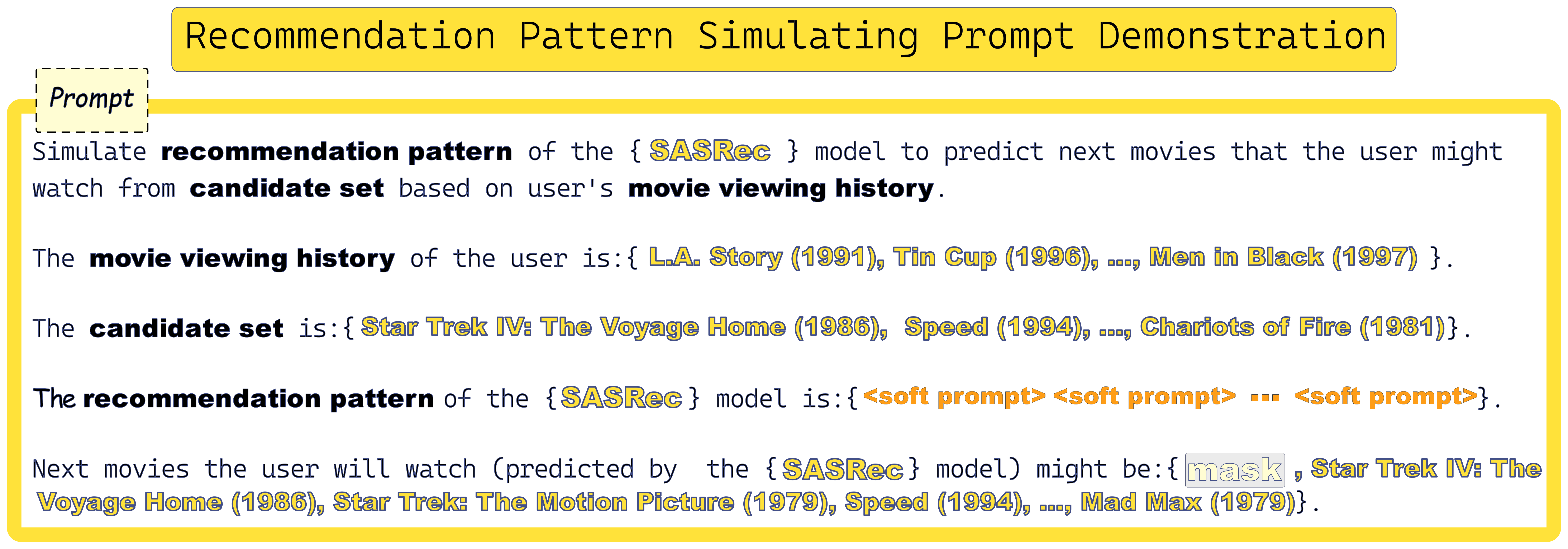}
  \caption{Demonstration of the prompt for \textit{Recommendation Pattern Simulating}.}
  \label{Figure 5}
  \vspace{-2mm}
\end{figure}

Specifically, we will let LLMs simulate the recommendation patterns of conventional SR models as closely as possible and let LLMs predict the recommendation results of conventional SR models (rather than the ground truth) based on the user interaction sequence. This process can be described as: given the user interaction sequence  $I_{1:n-1} = (I_1, I_2,..., I_{n-1})$ and providing the top $h$ recommended items $sr_{1:h} = (sr_1,sr_2,..., sr_{h})$ based on the conventional SR model's predicted probabilities for interaction sequence, we take the highest probability item $sr_1$ as the label $y^1$ for the \textit{Recommendation Pattern Simulating} task. Then, during the prediction of $y^1$, LLMs update the parameters of soft prompts, allowing LLMs to fit the results of the conventional SR model well. The prompt of the task is shown in Figure \ref{Figure 5}. Specifically, the loss function of \textit{Recommendation Pattern Simulating} (RPS) can be formulated as:

\vspace{-1.5mm}
\begin{equation}
\begin{aligned}
 L_{RPS} = \sum_{(x^1,y^1) \in D_1} -\log 
(P_{\Phi_0+\Phi}(y^1 | x^1)), 
\end{aligned}
\vspace{-1mm}
\end{equation}

\noindent where $D_1=\{(x_i^1, y_i^1)\}_{i=1, ..., N}$ consists of the prompt and conventional SR models predicted items in the RPS step.

After obtaining the loss functions for \textit{Temporal Analysis} and \textit{Recommendation Pattern Simulating}, we will proceed to update the parameters of soft prompts in a multi-task learning \cite{MTL} manner, allowing LLMs to learn from two target tasks simultaneously, thereby achieving the distillation of recommendation behavior patterns for conventional SR models. The learning objective can be defined as:
\begin{equation}
\begin{aligned}
\min_{\Phi} \{ \lambda L_{TA}+ (1 - \lambda) L_{RPS}\}, 
\end{aligned}
\end{equation}

\noindent where $\lambda$ and $1 - \lambda$ represent the weights of learning objectives of the two components, which are dynamically adjusted during training.

\subsection{LLMs-based Sequential Recommendation}
\label{LLMs-sr}

\begin{figure}[htbp]
\vspace{-2mm}
  \centering
  \includegraphics[width=0.95\linewidth]{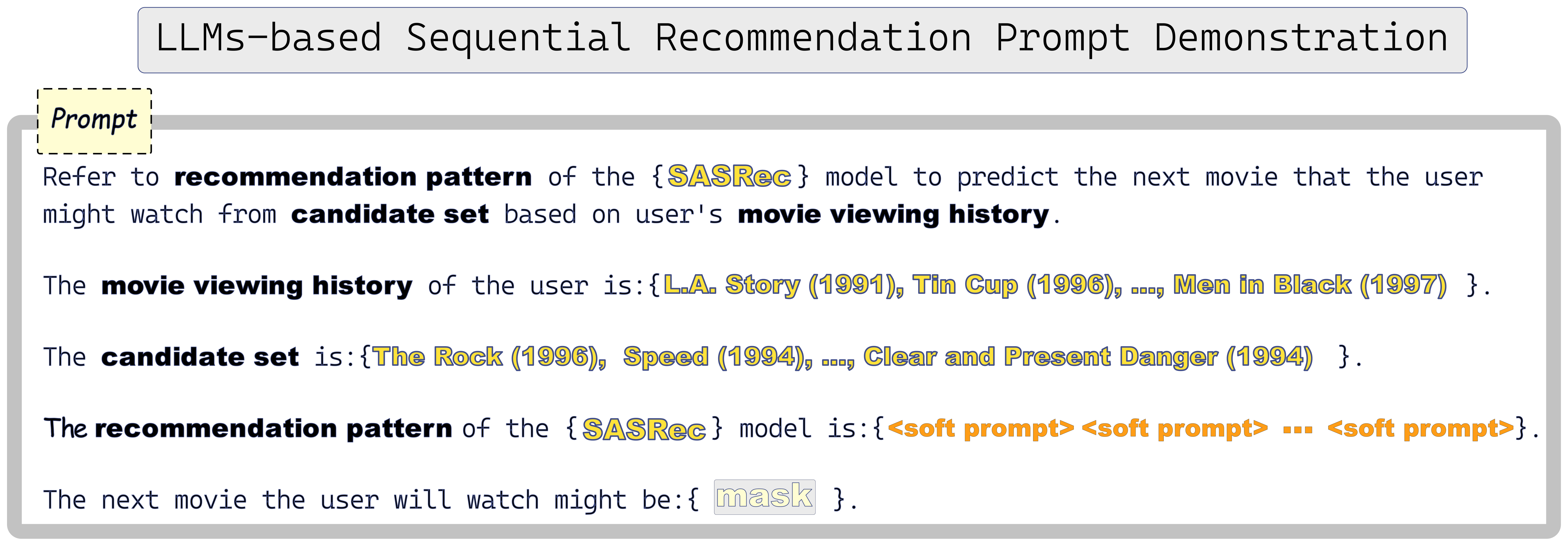}
  \caption{Demonstration of the prompt for \textit{LLMs-based Sequential Recommendation}. }
  \label{Figure 6}
  \vspace{-2mm}
\end{figure}

In the first stage (\textit{Distill Pattern from Conventional SR Models}), we successfully distilled the recommendation patterns from the conventional SR models. To enable LLMs to fully and effectively utilize the distilled auxiliary knowledge to achieve more accurate SR tasks, we proposed \textit{LLMs-based Sequential Recommendation}, the second stage of DELRec. In this stage, we insert soft prompts learned in the first stage into the prompt constructed in section \ref{prompt_construction} and freeze these parameters. Then, we fine-tune LLMs using ground truth.\\[0.25em]
\textbf{Insert Soft Prompts}.
In previous research, to enable LLMs to utilize auxiliary information from conventional SR models (such as item embeddings), people often used projectors ($e.g.,$ MLP, Tiny Transformers) to map the embeddings into the language space of LLMs. However, this approach often suffers from poorly designed projectors, which may fail to fully convey the information embedded in original embeddings to LLMs or limit their generalization capabilities. Therefore, distilled soft prompts will be inserted into the prompt we have constructed for the second stage, thereby overcoming previously mentioned issues. The prompt is shown in Figure \ref{Figure 6}.

Specifically, we incorporate the learned soft prompts $sp_{1:k} = (sp_1,sp_2,...,sp_{k})$ into the prompt $P$. In other words, we use the distilled recommendation behavior patterns from the conventional SR models as context for LLMs to predict the target $I_{n}$:
\begin{equation}
\begin{aligned}
I_{n} = LLM(P(sp_{1:k})) 
\end{aligned}
\end{equation}

\noindent where $P(\cdot)$ represents the process of inserting learned soft prompts $sp_{1:k}$ into constructed prompt $P$, and $LLM(\cdot)$ indicates that LLMs utilize the prompt to perform SR tasks.\\[0.25em]
\textbf{LLMs Fine-tuning}.
However, the distilled soft prompts may not be ideal for the language of fixed hard prompts, and these distilled soft prompts may contain noise. Therefore, we need to fine-tune LLMs' parameters to bridge the semantic gap between the two (distilled soft prompts and hard prompts) and guide LLMs to use the distilled soft prompts more as a reference, thus fully utilizing these distilled auxiliary information. Formally, given a user interaction sequence $I_{1:n-1} = (I_1,I_2,...,I_{n-1})$, where the next item $I_{n}$ the user will interact with is the label $\overline{y}$. In the learning process, when using soft prompts as auxiliary information to guide LLMs in predicting label $\overline{y}$, we freeze the parameters of soft prompts \(\Phi\) and fine-tune the LLMs using PEFT (AdaLora). Formally, the learning objectives of the LLMs-based Sequential Recommendation (LSR) can be described as follows:
\begin{equation}
\begin{aligned}
\min_{\Theta} \{ L_{LSR} = \sum_{(\overline{x},\overline{y}) \in \overline{D}} -\log 
(P_{\Phi_0+\Delta\Phi_0(\Theta)+\Phi}(\overline{y} | \overline{x})) \}, 
\end{aligned}
\end{equation}

\noindent where $\overline{D}=\{(\overline{x_i}, \overline{y_i})\}_{i=1, ...,N}$ contains the prompt $\overline{x}$ and the anticipated ground truth $\overline{y}$.

\section{EXPERIMENTS}
In this section, we assess the performance of our proposed framework, DELRec, on four real-world datasets. We compare it against various baselines, including conventional SR models and LLMs-based models. Our experiments are designed to answer the following research questions:\\[0.25em]
\textbf{RQ1}: Whether the proposed framework outperforms baseline methods, including the conventional SR models and other LLMs-based models, for SR? \\[0.25em]
\textbf{RQ2}: Are our proposed DELRec able to learn meaningful recommendation behavior patterns or information?\\[0.25em]
\textbf{RQ3}: How can key components affect our proposed method? Specifically, how is the efficacy of proposed \textit{Temporal Analysis} and \textit{Recommendation Pattern Simulating}?\\[0.25em]
\textbf{RQ4}: How do hyperparameters influence DELRec? How do the size and sparsity of datasets affect the performance of DELRec?\\[0.25em]
\textbf{RQ5}: What are DELRec's memory footprint, inference time and real-time response capability? Does DELRec face challenges with cold start issue?

\subsection{Setup}
\subsubsection{Datasets} 
We evaluate the proposed DELRec and baseline methods on four real-world datasets, namely MovieLens-100K, Beauty, Steam, and Home \& Kitchen, which are commonly used in the SR field. We show the detailed statistics of the datasets in Table \ref{table1}:\\[0.25em]
\textbf{MovieLens-100K}\footnote{\url{https://grouplens.org/datasets/movielens/100k/}}is a commonly used movie recommendation dataset that includes ratings given by users to movies and the titles of those movies.\\[0.25em]
\textbf{Steam}\footnote{\url{https://huggingface.co/datasets/joyliao7777/LLaRA/tree/main/steam}}not only contains user reviews of video games on the Steam Store, but also covers a variety of game titles.\\[0.25em] \textbf{Beauty}\footnote{\url{https://github.com/RUCAIBox/RecSysDatasets}}contains user feedback on beauty items from Amazon.\\[0.25em]
\textbf{Home \& Kitchen}\textsuperscript{3} is a dataset of user feedback containing various items needed for home and kitchen from Amazon website. This dataset contains millions of users and items to test the scalability of DELRec in real-world loads.

For all datasets, we follow \cite{SASRec} in treating users' implicit feedback as interactions between users and items, and determine the sequence order of inputs based on timestamps. Since users with few interactions are very sparse, it is difficult to extract meaningful patterns or preferences from them. The behavior of these users may not be representative and can easily introduce noise. Therefore, we filter out users and items with fewer than 5 interactions. Meanwhile, we arrange them chronologically as \cite{llara} do, and divide the data into training, validation, and test sets in an 8:1:1 ratio. This division method ensures that interactions used for training do not appear in subsequent data, thereby avoiding any potential information leakage.

\begin{table}[h]
\vspace{-4mm}
 \centering
  \caption{Statistics of Datasets.}
  \label{table1}
\resizebox{0.8\linewidth}{!}{
\begin{tabular}{lrrrr}
\toprule
Dataset          & sequence  & item      & interaction & sparsity \\ \hline
MovieLens-100K   & 943       & 1682      & 100,000     & 93.70\%  \\
Steam            & 11,938    & 3,581     & 274,726     & 99.36\%  \\
Beauty           & 324,038   & 32,586    & 371,345     & 99.99\%  \\
Home \& Kitchen & 9,767,606 & 1,286,050 & 21,928,568  & 99.99\%  \\ \bottomrule
\end{tabular}
}
\vspace{-2mm}
\end{table}

\subsubsection{Baselines} 
To demonstrate the effectiveness of our DELRec framework, we use two types of baseline models, namely conventional SR models and LLMs-based models. For conventional SR models, we will select the deep learning models that are more widely used and effective in the field of SR. We adopt three models that are often regarded as baselines:\\[0.25em]
\textbf{GRU4Rec}: GRU4Rec is based on RNN, which focuses on capturing temporal dependencies in user behavior sequences. It processes sequence data by GRU, which can effectively predict the users' next interaction.\\[0.25em]
\textbf{Caser}: Caser based on CNN predicts users' next interaction by capturing local and global patterns in behavior sequence.\\[0.25em]
\textbf{SASRec}: SASRec is an SR model based on a transformer, which aims to predict users' interaction by capturing patterns in the sequence of user behaviors. It uses the self-attention mechanism to model users' long-term and short-term preferences.

For the LLMs-based models, we will select three different open-source LLMs and the best-performing LLMs-based SR models (for a fair comparison, we will replace their LLM backbone with \textbf{Flan-T5-XL} \cite{flan_t5}) from three different paradigms described in Section \ref{three_para}:\\[0.25em]
\textbf{Bert-Large}: Bert-Large is a milestone LLM capable of performing Masked language modeling (MLM) tasks.\\[0.25em]
\textbf{Flan-T5-Large}: Flan-T5-Large (700M) are well-known open-source LLMs with an encoder-decoder structure.\\[0.25em]
\textbf{Flan-T5-XL}: Structurally similar to Flan-T5-Large, but due to the larger model size of Flan-T5-XL (3B), Flan-T5-XL performs better on complex tasks.\\[0.25em]
\textbf{RecRanker}: RecRanker cleverly samples items and users and inputs the results of conventional recommendation models into the prompt. It belongs to the first paradigm.\\[0.25em]
\textbf{LLMSEQPROMPT}: LLMSEQPROMPT enhances the accuracy of SR by integrating domain knowledge into prompts, fine-tuning LLMs, and then it generates recommendations based on prompts. It belongs to the first paradigm.\\[0.25em]
\textbf{LLM-TRSR}: LLM-TRSR analyzes users' historical interactions, generates recurrent summaries, and builds prompts that include users' preferences and candidates, then it uses prompts to fine-tune LLMs. It belongs to the first paradigm.\\[0.25em]
\textbf{LLaRA}: LLaRA inserts the embedding of items encoded by the conventional SR model into the prompt. It belongs to the second paradigm.\\[0.25em]
\textbf{LLM2BERT4Rec}: LLM2BERT4Rec uses LLMs embeddings to initialize the item embeddings in BERT4Rec, and then changes the dimension of the embeddings by using PCA as a projector before initialization. Then, it uses these embeddings to fine-tune BERT4Rec. It belongs to the second paradigm.\\[0.25em]
\textbf{LlamaRec}: LlamaRec recalls items with conventional model embeddings and uses a verbalizer to convert the output logits of LLMs into candidate probabilities. It belongs to the third paradigm.\\[0.25em]
\textbf{LLMSEQSIM}: LLMSEQSIM obtains item embeddings by using LLMs, then calculates session embeddings with item embeddings. It finally calculates similarity and recommends the most similar item to the user. It belongs to the third paradigm.\\[0.25em]
\textbf{KDA$_{\textbf{LRD}}$}: KDA$_{LRD}$ uses KDA \cite{KDA} as backbone, where KDA employs a Fourier-based temporal evolution module to track the dynamic changes in item relations over time. Then, it enhances KDA with LRD which was introduced in section \ref{KDA_label}. It belongs to the third paradigm.

We divide the proposed DELRec into three types, with Caser, GRU4Rec, and SASRec serving as the conventional SR model backbones in this framework. And we choose \textbf{Flan-T5-XL} as our LLMs backbone, because the main task in the framework we propose is Masked Language Modeling (MLM). We will also use \textbf{Flan-T5-Large} to perform ablation experiments. It's worth noting that the LLM backbones of our proposed framework can also use open-source Decoder-Only structured LLMs, such as Llama2 \cite{llama}, and are not constrained by the types of LLMs.

\vspace{-1mm}

\begin{table*}[h]
\centering
\caption{Overall Performance of DELRec on four datasets. The best-performing method in each column is boldfaced, and the second-best method (if not DELRec) in each column is underlined. The superscript $*$  and $**$ indicate \textit{p} $\leq$ 0.01 and \textit{p} $\leq$ 0.05 for the paired t-test of DELRec vs. conventional SR model backbone.}
\label{table2}
\resizebox{0.75\textwidth}{!}{
\begin{tabular}{llllllllllll}
\toprule
\multicolumn{1}{c}{} &
  \multicolumn{1}{r}{\textit{}} &
  \multicolumn{5}{c}{MovieLens-100K} &
  \multicolumn{5}{c}{Steam} \\ \hline
\multicolumn{1}{c}{} &
   &
  HR@1 &
  HR@5 &
  NDCG@5 &
  HR@10 &
  NDCG@10 &
  HR@1 &
  HR@5 &
  NDCG@5 &
  HR@10 &
  NDCG@10 \\ \hline
\multicolumn{1}{c}{} &
  Caser &
  0.3250 &
  0.6640 &
  0.5034 &
  0.7713 &
  0.5564 &
  0.3767 &
  0.6680 &
  0.5117 &
  0.7856 &
  0.5618 \\
\multicolumn{1}{l}{Conventional} &
  GRU4Rec &
  0.3162 &
  0.6495 &
  0.4824 &
  0.7568 &
  0.5430 &
  0.3786 &
  0.6835 &
  \underline{0.5241} &
  0.7915 &
  0.5672 \\
 &
  SASRec &
 \underline{0.3341} &
  0.6704 &
  \underline{0.5183} &
  0.8062 &
  0.5727 &
  0.3852 &
  \underline{0.6977} &
  \textbf{0.5305} &
  0.8337&
  0.5943 \\ \hline
 &
  Bert-Large &
  0.0053 &
  0.0204 &
  0.0126 &
  0.0404 &
  0.0236 &
  0.0081 &
  0.0424 &
  0.0349 &
  0.1367 &
  0.0587 \\
 &
  Flan-T5-Large &
  0.0375 &
  0.0703 &
  0.0509 &
  0.1529 &
  0.1076 &
  0.0540 &
  0.0893 &
  0.0586 &
  0.1938 &
  0.1237 \\
\multicolumn{1}{c}{} &
  Flan-T5-XL &
  0.0938 &
  0.2441 &
  0.1711 &
  0.3437 &
  0.2121 &
  0.1033 &
  0.2857 &
  0.1826 &
  0.4165 &
  0.2474 \\
 &
  LlamaRec &
  0.2870 &
  0.5873 &
  0.4253 &
  0.7183 &
  0.5387 &
  0.3511 &
  0.6278 &
  0.4522 &
  0.7691 &
  0.5489 \\
\multicolumn{1}{c}{} &
  RecRanker &
  0.3246 &
  0.6492 &
  0.4694 &
  0.7570 &
  0.5464 &
  0.3724 &
  0.6637 &
  0.4743 &
  0.7823 &
  0.5585 \\
\multicolumn{1}{l}{LLMs-based} &
  LLaRA &
  0.3523 &
  \underline{0.6753} &
  0.5016 &
  0.8089 &
  0.5632 &
  0.3855 &
  0.6811 &
  0.5081 &
  0.8461 &
  0.5943 \\
 &
  LLMSEQPROMPT &
  0.2702 &
  0.5681 &
  0.4152 &
  0.6921 &
  0.5332 &
  0.3597 &
  0.6439 &
  0.4546 &
  0.7740 &
  0.5538 \\
 &
  LLM2BERT4Rec &
  0.3365 &
  0.6507 &
  0.4743 &
  0.7615 &
  0.5518 &
  0.3825 &
  0.6742 &
  0.4886 &
  0.8163 &
  0.5706 \\
 &
  LLMSEQSIM &
  0.3397 &
  0.6532 &
  0.4728 &
  0.7690 &
  0.5521 &
  0.3848 &
  0.6895 &
  0.4937 &
  0.8171 &
  0.5711 \\
 &
  LLM-TRSR &
  0.3169 &
  0.6356 &
  0.4637 &
  0.7533 &
  0.5476 &
  0.3691 &
  0.6582 &
  0.4718 &
  0.7788 &
  0.5546 \\
 &
  KDA$_{LRD}$ &
  0.3618 &
  0.6734 &
  0.5089 &
  \underline{0.8185} &
  \underline{0.5824} &
  \underline{0.4028} &
  0.6904 &
  0.5138 &
  \underline{0.8522} &
  \underline{0.6077} \\ \hline
 &
  DELRec (Caser) &
  0.3664\textsuperscript{*} &
  0.6804\textsuperscript{**} &
  0.5108\textsuperscript{**} &
  0.8215\textsuperscript{*} &
  0.5843\textsuperscript{*} &
  0.4157\textsuperscript{**} &
  0.6946\textsuperscript{**} &
  0.5116 &
  0.8604\textsuperscript{*} &
  0.6108\textsuperscript{*} \\
\multicolumn{1}{l}{Ours} &
  DELRec (GRU4Rec) &
  0.3635\textsuperscript{*} &
  0.6722\textsuperscript{*} &
  0.5062\textsuperscript{*} &
  0.8137\textsuperscript{*} &
  0.5807\textsuperscript{*} &
  0.4296\textsuperscript{*} &
  0.7099\textsuperscript{**} &
  0.5192 &
  0.8679\textsuperscript{*} &
  0.6154\textsuperscript{*} \\
 &
  DELRec (SASRec) &
  \textbf{0.3701}\textsuperscript{*} &
  \textbf{0.6919}\textsuperscript{*} &
  \textbf{0.5237}\textsuperscript{*} &
  \textbf{0.8386}\textsuperscript{*} &
  \textbf{0.5950}\textsuperscript{*} &
  \textbf{0.4372}\textsuperscript{*} &
  \textbf{0.7285}\textsuperscript{*} &
  0.5286 &
  \textbf{0.8863}\textsuperscript{*} &
  \textbf{0.6203}\textsuperscript{*} \\ \bottomrule
\end{tabular}
}

\vspace{-2mm}

\resizebox{0.75\textwidth}{!}{
\begin{tabular}{llllllllllll}
\\ \\  \toprule
\multicolumn{1}{c}{}     & \multicolumn{1}{r}{\textit{}} & \multicolumn{5}{c}{Beauty}                        & \multicolumn{5}{c}{Home \& Kitchen}         \\ \hline
\multicolumn{1}{c}{}      &                               & HR@1   & HR@5   & NDCG@5       & HR@10  & NDCG@10 & HR@1   & HR@5   & NDCG@5 & HR@10  & NDCG@10 \\ \hline
\multicolumn{1}{c}{}             & Caser                         & 0.2241 & 0.4387 & 0.4073       & 0.4664 & 0.4257  & 0.1057 & 0.1597 & 0.1198 & 0.2184 & 0.1536  \\
\multicolumn{1}{l}{Conventional} & GRU4Rec                       & 0.2369 & 0.4544 & 0.4109       & 0.4755 & 0.4283  & 0.1091 & 0.1636 & 0.1259 & 0.2239 & 0.1581  \\
                                 & SASRec                        & 0.2573 & 0.4629 & 0.4212       & 0.5044 & 0.4368  & 0.1146 & 0.1779 & 0.1365 & 0.2427 & 0.1685  \\ \hline
                                 & Bert-Large                    & 0.0016 & 0.0154 & 0.0082       & 0.0327 & 0.011   & 0.0004 & 0.0013 & 0.0008 & 0.0057 & 0.0031  \\
                                 & Flan-T5-Large                 & 0.0195 & 0.0346 & 0.0265       & 0.1055 & 0.087   & 0.0086 & 0.0106 & 0.0066 & 0.0342 & 0.0135  \\
\multicolumn{1}{c}{}             & Flan-T5-XL                    & 0.0652 & 0.2071 & 0.1402       & 0.2709 & 0.1931  & 0.0257 & 0.0486 & 0.0301 & 0.0728 & 0.0427  \\
                                 & LlamaRec                      & 0.2361 & 0.4318 & 0.3879       & 0.4618 & 0.4065  & 0.1082 & 0.1618 & 0.1244 & 0.2215 & 0.1574  \\
\multicolumn{1}{c}{}             & RecRanker                     & 0.2470 & 0.4443 & 0.4051       & 0.4686 & 0.4274  & 0.1228 & 0.1811 & 0.1386 & 0.2570 & 0.1713  \\
\multicolumn{1}{l}{LLMs-based}   & LLaRA                         & 0.2752 & 0.4579 & 0.4173       & 0.4886 & 0.4310  & 0.1545 & 0.2046 & 0.1487 & 0.2892 & 0.1902  \\
                                 & LLMSEQPROMPT                  & 0.2390 & 0.4364 & 0.3981       & 0.4627 & 0.4183  & 0.1139 & 0.1752 & 0.1306 & 0.2366 & 0.1604  \\
                                 & LLM2BERT4Rec                  & 0.2575 & 0.4662 & 0.4223       & 0.5132 & 0.4392  & 0.1393 & 0.1870 & 0.1412 & 0.2743 & 0.1819  \\
                                 & LLMSEQSIM                     & 0.2739 & 0.4724 & \underline{0.4267} & 0.5268 & 0.4436  & 0.1481 & 0.1964 & 0.1452 & 0.2825 & 0.1876  \\
                                 & LLM-TRSR                      & 0.2482 & 0.4593 & 0.4195       & 0.4953 & 0.4342  & 0.1265 & 0.1825 & 0.1383 & 0.2661 & 0.1758  \\
 &
  KDA$_{LRD}$ &
  \underline{0.2856} &
  \underline{0.4886} &
  0.4261 &
  \underline{0.5366} &
  \underline{0.4485} &
  \underline{0.1684} &
  \underline{0.2207} &
  \underline{0.1523} &
  \underline{0.2957} &
  \underline{0.1945} \\ \hline
                                 & DELRec (Caser)                & 0.2819\textsuperscript{*} & 0.4850\textsuperscript{*} & 0.4266\textsuperscript{*}       & 0.5345\textsuperscript{*} & 0.4479\textsuperscript{*}  & 0.1707\textsuperscript{*} & 0.2256\textsuperscript{*} & 0.1564\textsuperscript{*} & 0.3025\textsuperscript{*} & 0.1987\textsuperscript{*}  \\
\multicolumn{1}{l}{Ours}         & DELRec (GRU4Rec)              & 0.3083\textsuperscript{*} & 0.4929\textsuperscript{*} & 0.4283\textsuperscript{*}       & 0.5519\textsuperscript{*} & 0.4576\textsuperscript{*}  & 0.1729\textsuperscript{*} & 0.2293\textsuperscript{*} & 0.1572\textsuperscript{*} & 0.3079\textsuperscript{*} & 0.2013\textsuperscript{*}  \\
 &
  DELRec (SASRec) &
  \textbf{0.3177}\textsuperscript{*} &
  \textbf{0.5013}\textsuperscript{*} &
  \textbf{0.4445}\textsuperscript{*} &
  \textbf{0.5730}\textsuperscript{*} &
  \textbf{0.4683}\textsuperscript{*} &
  \textbf{0.1836}\textsuperscript{*} &
  \textbf{0.2458}\textsuperscript{*} &
  \textbf{0.1639}\textsuperscript{*} &
  \textbf{0.3155}\textsuperscript{*}&
  \textbf{0.2064}\textsuperscript{*} \\ \bottomrule
\end{tabular}
}
\vspace{-5mm}
\end{table*}

\subsubsection{Implementation Details}
We will introduce our implementation details in two parts: conventional SR models part and LLMs-based SR model part. First, for the three conventional SR models used as backbones, the implementation details will be respectively modified:\\[0.25em]
\textbf{SASRec}: For the training of \textbf{SASRec}, we use the Adam optimizer and we set the embedding size as 100, and we also use two self-attention blocks with a learning rate of 1e-3, a dropout rate of 0.5 and a batch size of 128.\\[0.25em]
\textbf{Caser}: For the training of \textbf{Caser}, the set is similar to \textbf{SASRec}. We use the Adam optimizer and set the number of horizontal filters as 16. And we also set the embedding size as 100 with a learning rate of 1e-3, a dropout rate of 0.4 and a batch size of 128.\\[0.25em]
\textbf{GRU4Rec}: For for the training of \textbf{GRU4Rec}, we use the Adagrad optimizer. And we set the embedding size as 64 with a learning rate of 0.01, a dropout rate of 0.3 and a batch size of 50.

Secondly, for the parts of DELRec that use LLMs, the implementation details will be divided into two parts: the first stage of DELRec and the second stage of DELRec:\\[0.25em]
\textbf{The First Stage}: For the first stage of DELRec, we set the length of user interaction sequences $n$ as 10, which means setting the most recent 10 interactions as the user interaction sequence in order and padding sequences less than 10. we designate the number of user candidate items $m$ to be 15. In other words, these 15 candidate items are composed of one correct value for prediction and fourteen other randomly selected items. Regarding the number of examples $\alpha$ in the ICL of \textit{Temporal Analysis}, we have chosen $\alpha$ as 4 for MovieLens-100K and Beauty based on \cite{improve_awareness}, and we have selected $\alpha$ as 6 for Steam and Home \& Kitchen. For the first stage of DELRec, we use the Lion \cite{lion} optimizer, with a learning rate of 5e-3 and weight decay of 1e-5 and run on 10 Nvidia 3090 GPUs. \\[0.25em]
\textbf{The Seconde Stage}: For the second stage of DELRec, we use the same values of $n$ and $m$ as in the first stage, and also use AdaLoRA and Lion optimizer, with a learning rate of 1e-4 and weight decay of 1e-6 and run on 10 Nvidia 3090 GPUs. The code is at \url{https://github.com/haoge6660101/DELRec_hao}.

\subsubsection{Evaluation Metrics} 
For ranking evaluation, we use top-$k$ Hit Rate (HR) and Normalized Discounted Cumulative Gain (NDCG) as measurement metrics, specifically adopting HR$@$1, HR$@$5, HR$@$10, NDCG$@$5 and NDCG$@$10.

\subsection{Performance Comparison (\textbf{RQ1})}
Table \ref{table2} presents the performance of our method DELRec and various baselines under five evaluation metrics. Comparing DELRec with the aforementioned baseline models, we can derive the following observations:\\[0.25em]
\textbf{Conventional SR Models}: Combining the results from the table and t-test, we found that DELRec achieved significant performance improvements over all conventional SR model backbones on four datasets. It achieved almost the highest HR$@k$ and NDCG$@k$ scores among conventional SR models that only rely on user interactions. The key reason for this superior improvement is that DELRec effectively combines the information from conventional SR models with the powerful inference ability and rich world knowledge of LLMs, thereby providing more accurate recommendations.\\[0.25em]
\textbf{Open-source LLMs}: When comparing with some original open-source LLMs ($e.g.$, \textbf{BERT}, \textbf{Flan-T5}), it is evident that these baseline models not only underperform DELRec in recommendation tasks but also exhibit lower metrics compared to conventional SR models and other LLMs-based recommendation methods. This discrepancy is because while these LLMs possess strong generalization capabilities, they lack domain-specific knowledge and understanding of recommendation patterns, which hinders their performance in recommendation tasks. Therefore, it is crucial to provide appropriate auxiliary information to adapt LLMs to specific recommendation tasks.\\[0.25em]
\textbf{LLMs-based SR Models}: When considering the other LLMs-based methods we have chosen, the reasons for DELRec's superior performance can be analyzed from several perspectives. Firstly, some methods directly provide recommendation results extracted from conventional recommendation models to prompt LLMs ($e.g.,$ \textbf{RecRanker}, \textbf{LLMSEQPROMPT}, \textbf{LLM-TRSR}). For these methods, two main issues arise: 1) textual information often falls short in accurately and comprehensively describing the specific recommendation behavior patterns of conventional SR models; 2) without guiding information on users' past behavior, LLMs are limited to making decisions based solely on the provided results. Secondly, some methods provide user or item embeddings through dimension transformation to LLMs ($e.g.,$ \textbf{LLaRA}, \textbf{LLM2BERT4Rec}). The main reason that their performance is inferior to DELRec is: although those original embeddings contain information from conventional SR models, the dimension transformation of embeddings through poor designed projector leads to information loss, and ultimately, these embeddings do not align well with the semantic space of LLMs. Thirdly, while some methods ($e.g.$, \textbf{LlamaRec}, \textbf{LLMSEQSIM}, \textbf{KDA$_{\textbf{LRD}}$}) are able to filter out items information from conventional models for LLMs, they do not provide effective guidance information. There is still room for improvement in terms of providing guidance information for LLMs-based recommendations.\\[0.25em]
\textbf{Significance Test}: To ensure that DELRec's performance improvements are significant \cite{t_test} and not due to chance, we conducted significance test using t-test. We evaluated DELRec's performance compared to conventional SR model backbones based on results from multiple experiments. Results indicated that DELRec showed significant performance improvements over all conventional SR model backbones. These results are presented in Table \ref{table2}, where the superscript $*$ and $**$ indicate significance levels of \textit{p} $\leq$ 0.01 and \textit{p} $\leq$ 0.05, respectively.

\subsection{Ablation Studies \uppercase\expandafter{\romannumeral 1} (\textbf{RQ2})}
To address \textbf{RQ2}, we experimented on the soft prompts distilled in the first stage of DELRec and used \textbf{SASRec} as the backbone model. As these soft prompts do not correspond to natural language and are not easily interpretable by humans, we performed three transformations on a portion of the soft prompts to verify if they truly capture meaningful recommendation behavior patterns or information:\\[0.25em]
\textbf{\textit{w/o} SP (Soft Prompts)}: We removed the soft prompts section and the part of instruction that directs LLMs to refer to auxiliary information from the conventional SR models.\\[0.25em]
\textbf{\textit{w} MCP (Manual Construction Prompts)}: This transformation is similar to the general prompt where hard prompts are used to construct auxiliary information. For constructing auxiliary information, we attempted to describe the recommendation process of SASRec model in natural language and replaced the original soft prompts with it.\\[0.25em]
\textbf{\textit{w} USP (Untrained Soft Prompts)}: Soft prompts that have not undergone training in the first stage were initialized randomly and inserted into our prompt.

\begin{table*}[h]
\vspace{-7mm}
\caption{ Ablation studies \uppercase\expandafter{\romannumeral 1} for learned soft prompts on four datasets. The best-performing method in each column is boldfaced.}
\label{tabel3}
\centering
\vspace{-2mm}
\resizebox{0.75\textwidth}{!}{
\begin{tabular}{cllllllllll}
\toprule
        & \multicolumn{5}{c}{MovieLens-100K}          & \multicolumn{5}{c}{Steam}                   \\ \hline
        & HR@1   & HR@5   & NDCG@5 & HR@10  & NDCG@10 & HR@1   & HR@5   & NDCG@5 & HR@10  & NDCG@10 \\ \hline
\textit{w/o} SP & 0.3020 & 0.6211 & 0.4450 & 0.8057 & 0.5646  & 0.3426 & 0.6316 & 0.4785 & 0.8266 & 0.5731  \\
\textit{w} MCP  & 0.3106 & 0.6357 & 0.4603 & 0.8141 & 0.5712  & 0.3608 & 0.6620 & 0.4642 & 0.8531 & 0.5947  \\
\textit{w} USP  & 0.2752 & 0.5724 & 0.4296 & 0.7913 & 0.5259  & 0.2977 & 0.5964 & 0.4316 & 0.8136 & 0.5487  \\
Default &
  \textbf{0.3701} &
  \textbf{0.6919} &
  \textbf{0.5237} &
  \textbf{0.8386} &
  \textbf{0.5950} &
  \textbf{0.4372} &
  \textbf{0.7285} &
  \textbf{0.5286} &
  \textbf{0.8863} &
  \textbf{0.6203} \\ \bottomrule
\end{tabular}
}

\vspace{-3mm}

\resizebox{0.75\textwidth}{!}{
\begin{tabular}{lllllllllll}
\\ \\ \toprule
 & \multicolumn{5}{c}{Beauty}                  & \multicolumn{5}{c}{Home \& Kitchen}         \\ \hline
 & HR@1   & HR@5   & NDCG@5 & HR@10  & NDCG@10 & HR@1   & HR@5   & NDCG@5 & HR@10  & NDCG@10 \\ \hline
\textit{w/o} SP   & 0.2765 & 0.4762 & 0.4144 & 0.5365 & 0.4521  & 0.1539 & 0.2074 & 0.1356 & 0.2832 & 0.2483  \\
\textit{w} MCP    & 0.2698 & 0.4738 & 0.4088 & 0.5209 & 0.4463  & 0.1602 & 0.2215 & 0.1498 & 0.2968 & 0.2557  \\
\textit{w} USP    & 0.2384 & 0.4207 & 0.3764 & 0.4916 & 0.4210  & 0.1341 & 0.1936 & 0.1163 & 0.2541 & 0.2215  \\
Default &
  \textbf{0.3177} &
  \textbf{0.5013} &
  \textbf{0.4445} &
  \textbf{0.5730} &
  \textbf{0.4683} &
  \textbf{0.1836} &
  \textbf{0.2458} &
  \textbf{0.1639} &
  \textbf{0.3155} &
  \textbf{0.2604} \\ \bottomrule
\end{tabular}
}
\vspace{-3mm}
\end{table*}

\begin{table*}[h]
\vspace{-1mm}
\caption{Ablation studies \uppercase\expandafter{\romannumeral 2} on four datasets. The best-performing method in each column is boldfaced.}
\vspace{-2mm}
\label{tabel_ab}
\centering
\resizebox{0.75\textwidth}{!}{
\begin{tabular}{lllllllllll}
\toprule
       & \multicolumn{5}{c}{MovieLens-100K}          & \multicolumn{5}{c}{Steam}                   \\ \hline
        & HR@1   & HR@5   & NDCG@5 & HR@10  & NDCG@10 & HR@1   & HR@5   & NDCG@5 & HR@10  & NDCG@10 \\ \hline
\textit{w/o} DPSM        & 0.3020 & 0.6211 & 0.4450 & 0.8057 & 0.5646  & 0.3426 & 0.6316 & 0.4785 & 0.8266 & 0.5731  \\
\textit{w/o} LSR         & 0.2814 & 0.5843 & 0.4223 & 0.7845 & 0.5314  & 0.3235 & 0.6120 & 0.4526 & 0.8105 & 0.5744  \\
\textit{w/o} TA          & 0.3425 & 0.6617 & 0.4867 & 0.8231 & 0.5797  & 0.3710 & 0.6676 & 0.4945 & 0.8450 & 0.5909  \\
\textit{w/o} PRS         & 0.3379 & 0.6585 & 0.4805 & 0.8187 & 0.5715  & 0.3555 & 0.6435 & 0.4798 & 0.8313 & 0.5886  \\
\textit{w} UDPSM         & 0.3513 & 0.6726 & 0.5013 & 0.8324 & 0.5906  & 0.3974 & 0.6969 & 0.5071 & 0.8687 & 0.6142  \\
\textit{w} ULSR          & 0.3486 & 0.6678 & 0.4930 & 0.8276 & 0.5823  & 0.3836 & 0.6829 & 0.4983 & 0.8542 & 0.6074  \\
\textit{w} Flan-T5-Large & 0.2592 & 0.5446 & 0.4074 & 0.7639 & 0.5008  & 0.3018 & 0.5772 & 0.4279 & 0.7924 & 0.5687  \\
Default &
  \textbf{0.3701} &
  \textbf{0.6919} &
  \textbf{0.5237} &
  \textbf{0.8386} &
  \textbf{0.5950} &
  \textbf{0.4372} &
  \textbf{0.7285} &
  \textbf{0.5286} &
  \textbf{0.8863} &
  \textbf{0.6203} \\ \bottomrule
\end{tabular}
}

\vspace{-3mm}

\resizebox{0.75\textwidth}{!}{
\begin{tabular}{lllllllllll}
\\ \\ \toprule
       & \multicolumn{5}{c}{Beauty}                  & \multicolumn{5}{c}{Home \& Kitchen}         \\ \hline
        & HR@1   & HR@5   & NDCG@5 & HR@10  & NDCG@10 & HR@1   & HR@5   & NDCG@5 & HR@10  & NDCG@10 \\ \hline
\textit{w/o} DPSM        & 0.2765 & 0.4762 & 0.4144 & 0.5365 & 0.4521  & 0.1539 & 0.2074 & 0.1356 & 0.2832 & 0.2483  \\
\textit{w/o} LSR         & 0.2666 & 0.4683 & 0.4117 & 0.5287 & 0.4475  & 0.1491 & 0.2013 & 0.1280 & 0.2777 & 0.2425  \\
\textit{w/o} TA          & 0.2815 & 0.4785 & 0.4192 & 0.5396 & 0.4563  & 0.1564 & 0.2156 & 0.1432 & 0.2915 & 0.2490  \\
\textit{w/o} PRS         & 0.2903 & 0.4861 & 0.4294 & 0.5431 & 0.4587  & 0.1682 & 0.2234 & 0.1486 & 0.2967 & 0.2512  \\
\textit{w} UDPSM         & 0.3067 & 0.4927 & 0.4373 & 0.5582 & 0.4644  & 0.1738 & 0.2335 & 0.1612 & 0.3082 & 0.2588  \\
\textit{w} ULSR          & 0.2945 & 0.4886 & 0.4321 & 0.5506 & 0.4619  & 0.1704 & 0.2272 & 0.1566 & 0.3024 & 0.2561  \\
\textit{w} Flan-T5-Large & 0.2384 & 0.4362 & 0.3865 & 0.5013 & 0.4288  & 0.1235 & 0.1867 & 0.1139 & 0.2594 & 0.2373  \\
Default &
  \textbf{0.3177} &
  \textbf{0.5013} &
  \textbf{0.4445} &
  \textbf{0.5730} &
  \textbf{0.4683} &
  \textbf{0.1836} &
  \textbf{0.2458} &
  \textbf{0.1639} &
  \textbf{0.3155} &
  \textbf{0.2604} \\ \hline
\end{tabular}
}

\vspace{-5mm}
\end{table*}

Finally, the three transformed methods are compared with the complete DELRec after fine-tuning in the second stage. Table \ref{tabel3} shows the measurement metrics of DELRec under four different conditions. Based on our observations, we make the following inferences:

\begin{itemize}[leftmargin=*]
\item Soft prompts that have undergone our designed \textit{Distill Pattern from Conventional SR Models} approach surpass all three abovementioned methods. This indicates that our distillation method is able to effectively extract valuable recommendation patterns and information from conventional SR models for LLMs.

\item In methods that solely utilize pure hard prompts without soft prompts or manual construction, the Manual Construction method enhances LLMs by describing the recommendation patterns of the conventional SR model in nature language. But due to inaccuracies or insufficient information in these descriptions, the metrics of this method only show slight improvements compared to the No Soft Prompts method. 

\item Among the few baselines we selected, the Random Soft Prompts method performs poorly in metrics. This can be attributed to random soft prompts being scattered throughout the semantic space with no meaningful context, resulting in strong noise and providing little assistance or potentially misleading LLMs. 

\end{itemize}

\subsection{Ablation Studies \uppercase\expandafter{\romannumeral 2} (\textbf{RQ3})}
We will verify the impact of various components in DELRec on the framework through the following ablation experiments. The results are shown in Table \ref{tabel_ab}. We introduce the variants and analyze their effect respectively:\\[0.25em]
\textbf{\textit{w/o} DPSM (\textit{Distill Pattern from Conventional SR Models})}: By removing the process of distilling recommendation behavior patterns from conventional SR models in the first stage of DELRec, we observed a decline in performance. This is because LLMs lack auxiliary information from conventional SR models, which hinders their ability to effectively guide the recommendation process.\\[0.25em]
\textbf{\textit{w/o} LSR (\textit{LLMs-based Sequential Recommendation})}: After distillation in the first stage, we remove the fine-tuning process of LLMs in the second stage. This led to a decline in metrics and caused LLMs to exhibit a greater tendency toward favoring items predicted by conventional SR models. It can be attributed to using information extracted directly from conventional SR models introduces noise that may interfere with LLMs-based recommendations.\\[0.25em]
\textbf{\textit{w/o} TA (\textit{Temporal Analysis})}: Removing \textit{Temporal Analysis} during the first stage caused insufficient guidance for LLMs to simulate feature aggregation processes similar to conventional SR models. The reason is that distilled soft prompts lack temporal characteristics. \\[0.25em]
\textbf{\textit{w/o} RPS (\textit{Recommendation Pattern Simulating})}: Eliminating \textit{Recommendation Pattern Simulating} during the first stage shows a decline in metrics, and it becomes challenging for LLMs to effectively simulate overall recommendation behavior patterns exhibited by conventional SR models. It can be attributed to disrupting integration between prediction results of those of conventional SR models and LLMs.\\[0.25em]
\textbf{\textit{w} UDPSM (Updating both soft prompts and LLMs parameters in DPSM)}: We observed a decline in performance when updating both soft prompts and LLMs' parameters in DELRec's first stage, rather than just the soft prompts. This simultaneous updating prevents soft prompts from being updated independently, and this results in reducing information flow to the soft prompts and hindering SR pattern distillation.\\[0.25em]
\textbf{\textit{w} ULSR (Updating both soft prompts and LLMs parameters in LSR)}: We observed a decline in performance when updating both soft prompts' and LLMs' parameters in the second stage of DELRec, rather than only updating the LLMs. The reason is that updating soft prompt parameters in the second stage causes the loss of information and recommendation pattern distilled in the first stage, thereby resulting in those soft prompts unable to provide better reference for LLMs.\\[0.25em]
\textbf{\textit{w} Flan-T5-Large}: In addition to ablation experiments on DELRec components, we explored using \textbf{Flan-T5-Large} as a smaller-scale LLM backbone. Results indicated that both the size and capacity of LLMs impact DELRec's performance.

\begin{table*}[h]
\vspace{-8mm}
\centering
\caption{Dataset Sparsity Impact on three datasets. The best-performing method in each column is boldfaced, and the second-best method in each column is underlined. The sparsity of each dataset is indicated in parentheses next to the dataset's name.}
\vspace{-2mm}
\label{tabel_5}
\resizebox{1\textwidth}{!}{
\begin{tabular}{llllllllllllllll}
\toprule
\multicolumn{1}{r}{\textit{}} &
  \multicolumn{5}{c}{Beauty (99.99\%)} &
  \multicolumn{5}{c}{MovieLens-100K (93.70\%)} &
  \multicolumn{5}{c}{KuaiRec (83.72\%)} \\ \hline
 &
  HR@1 &
  HR@5 &
  NDCG@5 &
  HR@10 &
  NDCG@10 &
  HR@1 &
  HR@5 &
  NDCG@5 &
  HR@10 &
  NDCG@10 &
  HR@1 &
  HR@5 &
  NDCG@5 &
  HR@10 &
  NDCG@10 \\ \hline
SASRec &
  0.2573 &
  0.4629 &
  0.4212 &
  0.5044 &
  0.4368 &
  0.3341 &
  0.6704 &
  \underline{0.5183} &
  0.8062 &
  0.5727 &
  0.5379 &
  0.7934 &
  0.5794 &
  0.8737 &
  0.6659 \\
KDA$_{LRD}$ &
  \underline{0.2856} &
  \underline{0.4886} &
  \underline{0.4261} &
  \underline{0.5366} &
  \underline{0.4485} &
  \underline{0.3618} &
  \underline{0.6734} &
  0.5089 &
  \underline{0.8185} &
  \underline{0.5824} &
  \underline{0.5446} &
 \underline{0.8061} &
  \underline{0.5826} &
  \underline{0.8865} &
  \underline{0.6715} \\
DELRec &
  \textbf{0.3177} &
  \textbf{0.5013} &
  \textbf{0.4445} &
  \textbf{0.5730} &
  \textbf{0.4683} &
  \textbf{0.3701} &
  \textbf{0.6919} &
  \textbf{0.5237} &
  \textbf{0.8386} &
  \textbf{0.5950} &
  \textbf{0.5615} &
  \textbf{0.8243} &
  \textbf{0.5881} &
  \textbf{0.9001} &
  \textbf{0.6803}\\ \bottomrule
\end{tabular}
}
\vspace{-4mm}
\end{table*}

\vspace{-1mm}
\subsection{Dataset Impact and Hyperparameter Analysis (\textbf{RQ4})}
\begin{figure}[!ht]
  \vspace{-2mm}
  \centering
  \includegraphics[width=0.65\linewidth]{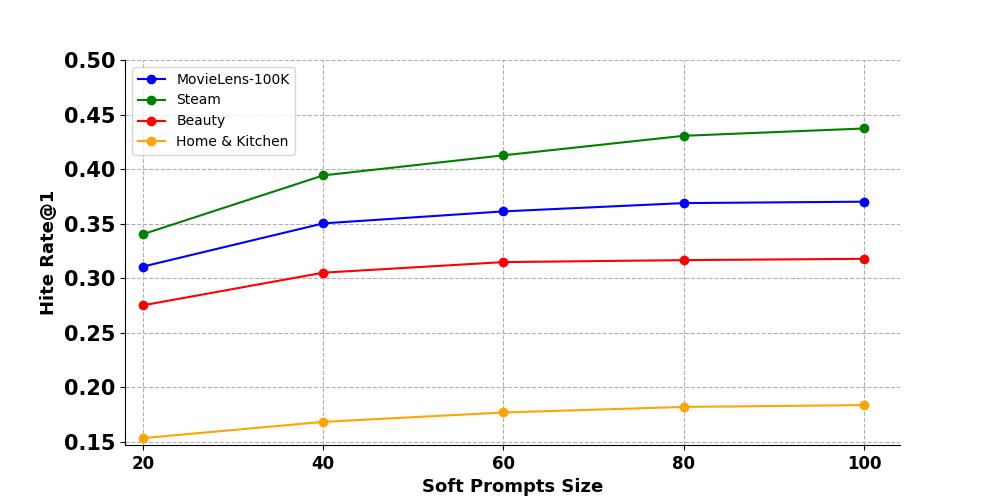}
  \vspace{-3mm}
  \caption{Performance comparison w.r.t different soft prompts size $k$ for training DELRec on the four datasets.}
  \label{fig soft}
  
  \centering
  \includegraphics[width=0.65\linewidth]{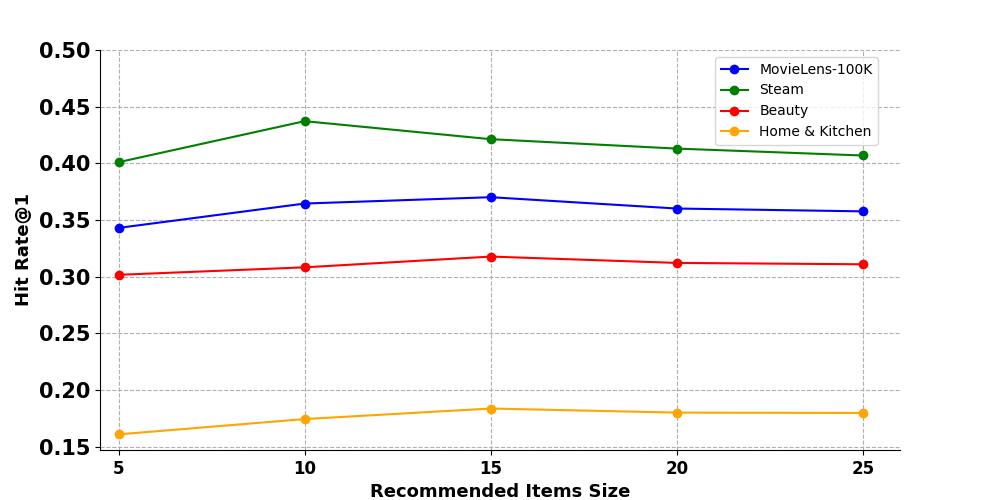}
  \vspace{-3mm}
  \caption{Performance comparison w.r.t different recommended items size $h$ for training DELRec on the four datasets.}
  \label{fig item}
  \vspace{-3mm}
\end{figure}

We experimented with the hyperparameters in DELRec, including the soft prompts size $k$ and the top $h$ recommended items from the conventional SR model ($e.g.$, \textbf{SASRec}). We used the HR@1 metric to display results, as it directly reflects the model's ability to recommend the most relevant item, crucial for user satisfaction in SR.\\[0.25em]
\textbf{Soft Prompts Size}: We examined the impact of soft prompts size $k$ on DELRec's performance. As shown in Figure \ref{fig soft}, DELRec's metrics initially improve with an increase in $k$. However, after reaching 80, the metrics level off. This suggests that while soft prompts enhance prompt information through LLMs' training, an excessive amount can introduce noise or lead to overfitting. Thus, beyond a certain size, additional soft prompts do not significantly improve overall performance.\\[0.25em]
\textbf{Recommended Items Size}: We investigated how the overall performance changes with varying sizes h of recommended items provided by conventional SR model during \textit{Recommendation Pattern Simulating}. Figure \ref{fig item} shows a relationship between $h$ and performance. Providing conventional SR model-recommended items helps LLMs understand recommendation patterns. However, too large recommended items size may mislead LLMs and result in excessively long prompts, which could potentially undermine LLMs' attention mechanism.

We also conducted experiments to analyze the impact of dataset size and sparsity on DELRec's performance.\\[0.25em]
\textbf{Dataset Size}: Regarding the impact of dataset size on DELRec's performance, the results in Table \ref{table2} show that DELRec's performance does not necessarily increase with the dataset size. This indicates that dataset size does not significantly affect DELRec's performance, demonstrating its scalability.\\[0.25em]
\textbf{Dataset Sparsity}: To better analyze the impact of dataset sparsity on DELRec's performance, we added a dataset called \textbf{KuaiRec}\footnote{\url{https://github.com/chongminggao/KuaiRec}}(a user viewing record dataset from Kuaishou app, containing 7,176 users, 10,728 items, and approximately 12,530,806 interactions, with about 83.72\% sparsity). We selected \textbf{KuaiRec} as the sparsest dataset among those we used to more clearly differentiate the sparsity levels of different datasets. Then we compared DELRec with \textbf{KDA$_{{LRD}}$} (a SOTA method for LLMs-based SR) and \textbf{SASRec} (a SOTA method for conventional SR) on three datasets: \textbf{MovieLens-100K}, \textbf{Beauty}, and \textbf{KuaiRec}. As shown in Table \ref{tabel_5}, as dataset sparsity increases, excessive noise makes it difficult for DELRec to accurately capture recommendation patterns and predict unobserved interactions, resulting in performance decline. Even though DELRec's performance declines as sparsity increases, it consistently outperforms other baselines across all datasets.

\subsection{Computational Efficiency Analysis (\textbf{RQ5})}
To answer the first question of \textbf{RQ5}, we tested the memory footprint and inference time of DELRec to verify its scalability under real-world loads:\\[0.25em]
\textbf{Memory Footprint}: The designed DELRec has around 3 billion parameters with its LLM backbone (\textbf{Flan-T5-XL}) and 0.2 million parameters for soft prompts, resulting in a model memory footprint of about 12 GB. When we use the largest dataset (\textbf{Home \& Kitchen}), the peak memory footprint during training is approximately 234 GB; during inferring, the peak memory footprint is approximately 68 GB.\\[0.25em]
\textbf{Inference Time \& Real-Time Response Capability}: Regarding the inference time of DELRec, we processed 1,000 requests using 10 Nvidia 3090 GPUs in batch. The total inference time was approximately 181.69\textit{s} (the average of ten calculations), with an average processing time of 0.182\textit{s} per request. In contrast, the LLM backbone (\textbf{Flan-T5-XL}) of DELRec, has an average inference time of 0.161\textit{s}, which is only 21\textit{ms} faster than DELRec. But DELRec significantly improves the accuracy of SR tasks and demonstrates good real-time response capability.\\[0.25em]
\textbf{Cold Start}: Regarding the second question of \textbf{RQ5}, it is noteworthy that DELRec does not exhibit significant shortcomings in cold start problem. This is because DELRec, an LLMs-based SR framework, has been pre-trained and accumulated rich world and recommendation domain knowledge through the first stage of distilled soft prompts. To verify this, we conducted a comparative experiment specifically for users with very few interactions (fewer than 3 interactions) on \textbf{Home \& Kitchen} to closely mimic real-world loads. The results show that, with only a tiny amount of interactions, DELRec's metrics (\textit{HR@1: 0.1082, HR@5: 0.1736, NDCG@5: 0.1344, HR@10: 0.2391, NDCG@10: 0.1605}) outperform \textbf{SASRec} (\textit{HR@1: 0.0954, HR@5: 0.1418, NDCG@5: 0.1180, HR@10: 0.1937, NDCG@10: 0.1495}) and are almost on par with \textbf{KDA$_{{LRD}}$} (\textit{HR@1: 0.1077, HR@5: 0.1759, NDCG@5: 0.1358, HR@10: 0.2383, NDCG@10: 0.1587}).

\subsection{Case Study}
\begin{figure}[h]
\vspace{-5mm}
  \centering
  \includegraphics[width=0.8\linewidth]{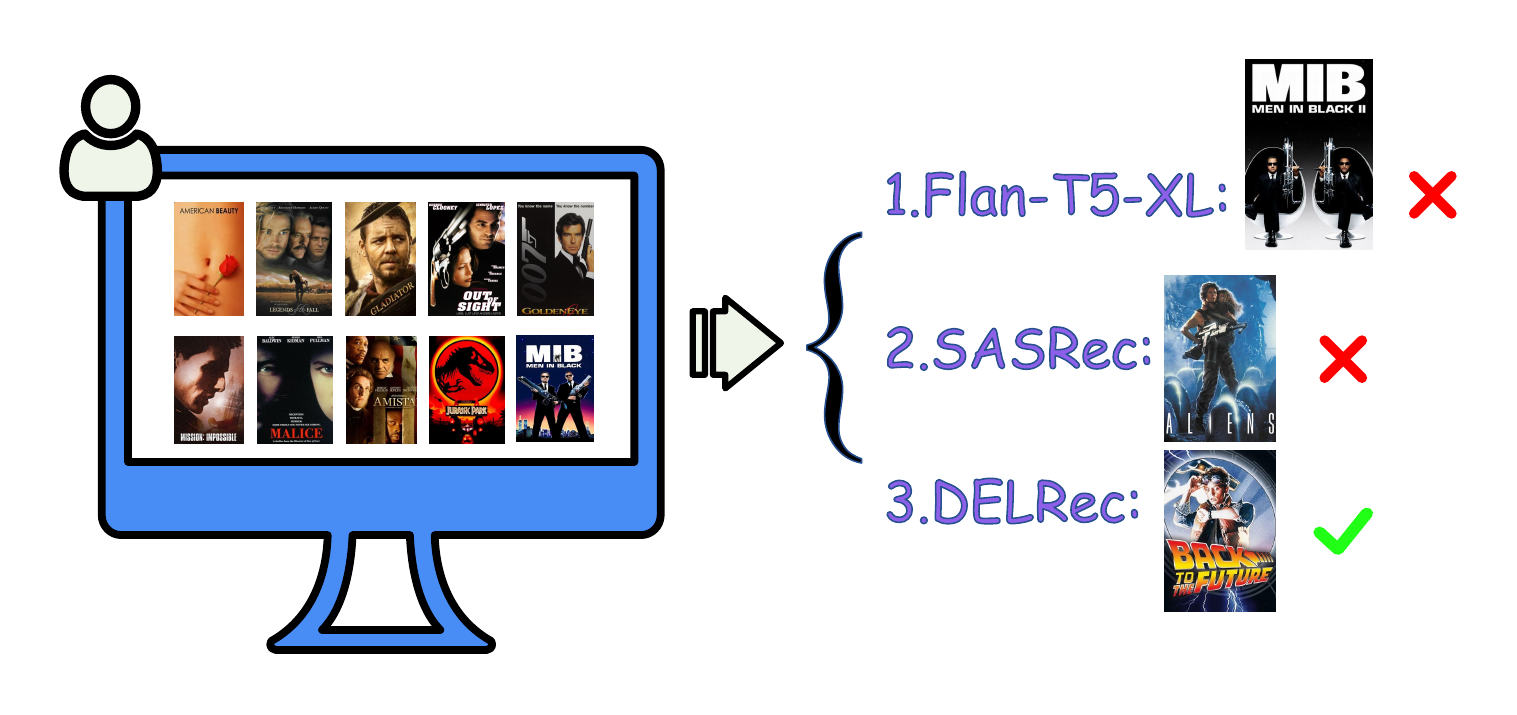}
  \vspace{-4mm}
  \caption{Case study comparison results of the effectiveness of three models in recommending movies.}
  \label{figure case}
  \vspace{-4mm}
\end{figure}

To investigate the effectiveness of integrating recommendation patterns from conventional SR models with LLMs' world knowledge, we conducted a comparative case study using \textbf{Flan-T5-XL}, \textbf{SASRec}, and DELRec. Here we choose a distinct example. For a user with a viewing history that includes "American Beauty (1999)", "Legends of the Fall (1994)", "Gladiator (2000)", "Out of Sight (1998)", "GoldenEye (1995)", "Mission: Impossible(1996)", "Malice (1993)", "Amistad (1997)", "Jurassic Park (1993)" and "Men in Black (1997)", we generated recommendations using the three models. As shown in Figure \ref{figure case}, \textbf{Flan-T5-XL} recommended "Men in Black II (2002)" based on the name of user's last watched movie. \textbf{SASRec}, considering user's recent viewing history, recommended the action/sci-fi film "Aliens (1986)", which aligns with the theme of "Men in Black". In contrast, DELRec combined conventional SR patterns with rich world knowledge and considered user's shift in preferences from drama/classic to action/sci-fi genres and recommended "Back to the Future (1985)", which was indeed user's next interaction.

\section{CONCLUSION}
This work introduces DELRec, a framework that enhances LLMs in SR tasks by extracting behavioral patterns from conventional SR models. Through proposed DPSM and LSR, DELRec reduces information loss and improves SR effectiveness. Extensive experiments on four real-world datasets validate our framework. DELRec offers a new approach for utilizing LLMs in complex SR tasks, capturing information and global context missed by conventional SR models.

\section{ACKNOWLEDGEMENT}
This work was supported by Shanghai Science and Technology Commission (No. 22YF1401100), Fundamental Research Funds for the Central Universities (No. 24D111201), National Science Fund for Young Scholars (No. 62202095), and the Open Project Program of the State Key Laboratory of CAD\&CG (Grant No. A241), Zhejiang University.

\vspace{12pt}

\bibliographystyle{IEEEtran}
\bibliography{sample}

\end{document}